\documentclass{article}

\usepackage{arxiv}

\usepackage[utf8]{inputenc} % allow utf-8 input
\usepackage[T1]{fontenc}    % use 8-bit T1 fonts
\usepackage{hyperref}       % hyperlinks
\usepackage{url}            % simple URL typesetting
\usepackage{booktabs}       % professional-quality tables
\usepackage{amsfonts}       % blackboard math symbols
\usepackage{nicefrac}       % compact symbols for 1/2, etc.
\usepackage{microtype}      % microtypography
\usepackage{lipsum}

\usepackage{graphicx}
\usepackage[table]{xcolor}% http://ctan.org/pkg/xcolor

\title{Understanding the impact of the alphabetical ordering of names in user interfaces: a gender bias analysis}

\author{
  Daniel~Sullivan \\
   Programa de P\'os Gradua\c{c}\~ao em Inform\'atica Aplicada\\
  Universidade de Fortaleza\\
  Fortaleza, Brazil \\
  \texttt{daniel.sullivan@edu.unifor.br} \\
   \And
 Carlos~Caminha \\
  Centro de Ci\^encias Tecnol\'ogicas\\
  Universidade de Fortaleza\\
  Fortaleza, Brazil \\
  \texttt{caminha@unifor.br} \\
 \And
 Victor~Dantas \\
  Programa de P\'os Gradua\c{c}\~ao em Inform\'atica Aplicada\\
  Universidade de Fortaleza\\
  Fortaleza, Brazil \\
  \texttt{victordantas2@gmail.com} \\
  \And
 Elizabeth~Furtado \\
  Programa de P\'os Gradua\c{c}\~ao em Inform\'atica Aplicada\\
  Universidade de Fortaleza\\
  Fortaleza, Brazil \\
  \texttt{elizabet@unifor.br} \\
    \And
 Vasco~Furtado \\
  Programa de P\'os Gradua\c{c}\~ao em Inform\'atica Aplicada\\
  Universidade de Fortaleza\\
  Fortaleza, Brazil \\
  \texttt{vasco@unifor.br} \\
   \And
 Virg\'ilio~Almeida \\
 Dept. of Computer Science\\
Universidade Federal de Minas Gerais\\
  Minas Gerais, Brazil \\
  \texttt{virgilio@dcc.ufmg.br} \\
}

\begin{document}
\maketitle

\begin{abstract}
Listing people alphabetically on an electronic output device is a traditional technique, since alphabetical order is easily perceived by users and facilitates access to information. However, this apparently harmless technique, especially when the list is ordered by first name, needs to be used with caution by designers and programmers. We show, via empirical data analysis, that when an interface displays people’s first name in alphabetical order in several pages/screens, each page/screen may have imbalances in respect to gender of its {\it Top-k} individuals. {\it k} represents the size of the list of names visualized first, which may be the number of names that fits in a screen page of a certain device. The research work was carried out with the analysis of actual datasets of names of five different countries. Each dataset has a person name and the frequency of adoption of the name in the country. Our analysis shows that, even though all countries have exhibit imbalance problems, the samples of individuals with Brazilian and Spanish first names are more prone to gender imbalance among their {\it Top-k} individuals. These results can be useful for designers and engineers to construct information systems that avoid gender bias induction.
\end{abstract}

% keywords can be removed
\keywords{Interface Systems \and Gender Bias \and Fairness Measure \and Alphabetical Order}

\section{Introduction}
In the context of Computer Science, an information system is considered biased when it systematically and unfairly discriminates against certain individuals or groups of individuals in favor of others \cite{friedman1996bias}. Bias in information systems is not a new problem being intimately related to the way information is displayed and visualized. Friedman and Nissenbaum \cite{friedman1996bias} described the controversial argument, from 40 years ago, between US airlines claiming that American and United Airlines, through its air reservation information systems, made available to travel agencies, implemented strategies that favored the two to the detriment of others \cite{shifrin1985justice}. The claim was that 90\% of bookings made by travel agents were from travel options that appeared on the first display screen (with no need to scroll down the page) \cite{mohd1990loophole}. American and United flights were more likely to be chosen because they appear first. Since then, rankings and orderings and the way they are displayed in devices have been investigated for their ability to induce bias in various domains \cite{carney2012first,einav2006s,feenberg2017s,haque2009positional,jacobs2015alphabetic}. In general, the problem of having a fair ranking so that the {\it Top-k} are exploited in an equanimous manner has been characterized \cite{zehlike2017fa}. 

Classical HCI research works have already shown that the more distant the item is from a user the less is the probability to be chosen by the users \cite{fitts1954information, pirolli2007information}.  Also, usability tests on scrolling user behavior show that Web users spend 80\% of their time looking at information above the page fold. Although users do scroll, they allocate only 20\% of their attention below the fold. It is natural to infer that items at the last screen page do not have the same chance of being picked than the first ones.

\begin{figure*}[t]
\centering
\includegraphics[width=1\linewidth]{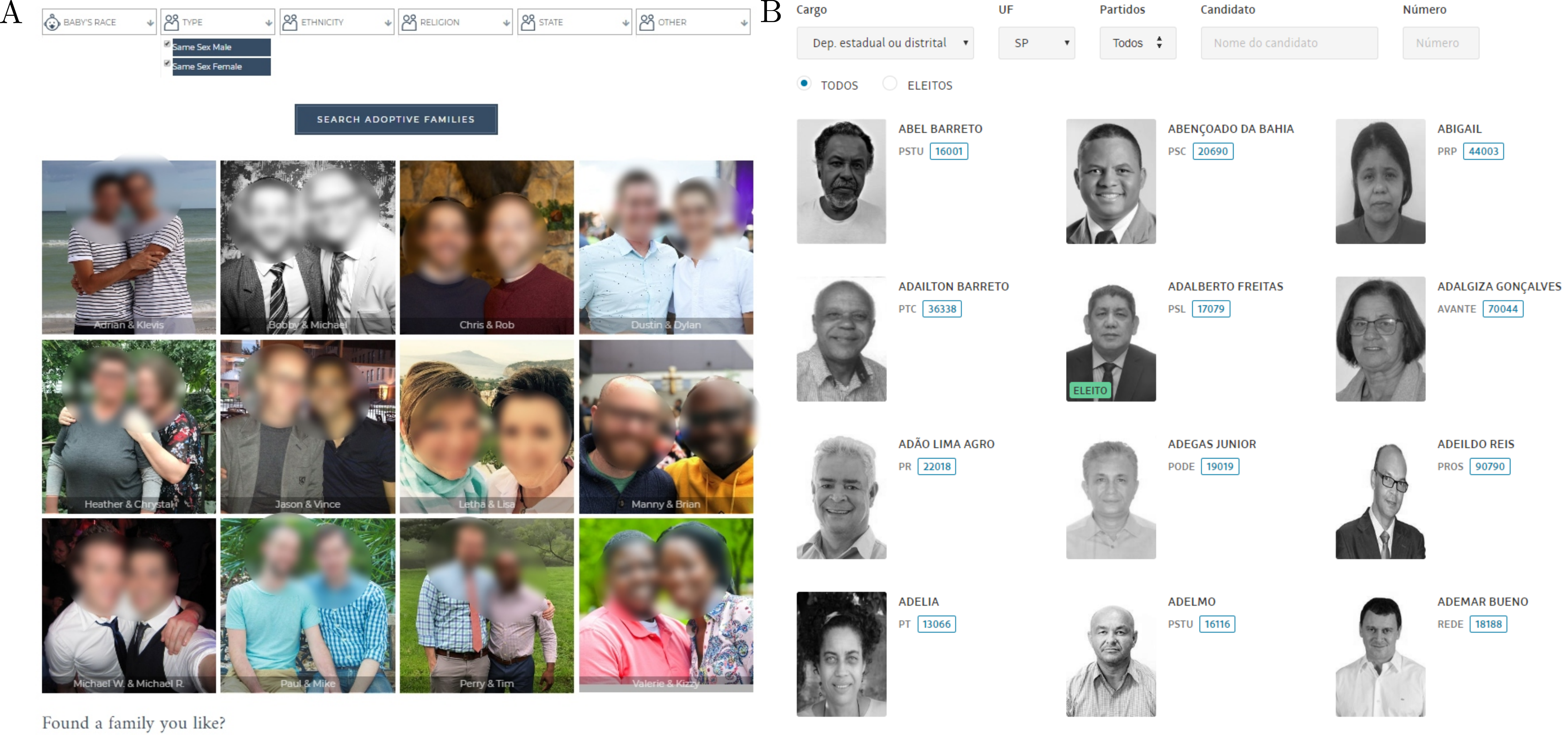}
\caption{
Examples of systems that support user decision making by choosing individuals alphabetically ordered by their first names. On the left is the interface of an adoption support system where the names of same-sex couples wishing to adopt are introduced to people who want to donate. The couples are alphabetically ordered by one of the partners name. On the right, a system to support the choice of candidates for the Brazilian Parliament. Candidates are grouped by the States of the Federation in alphabetical order}
\label{fig:exemplo-sistemas}
\end{figure*}

While some work has pointed to the risks of favoring those that appear first in list views, there is no major concern for HCI designers and software engineers in general when it comes to creating programs that display people's names in alphabetical order. In a quick survey on the Web, we found several examples of information systems in different languages that display a list of individuals in an alphabetical order of their first name. Figure \ref{fig:exemplo-sistemas} shows screen photos of two different information system interfaces, publicly available, in which an alphabetically ordered list of individuals is provided for a user decision making. In A, an adoption support system \cite{adoption-system} displays same-sex couples who share the  desire to adopt a child. The photo of the couples appears in several screen pages sorted alphabetically by the first name of one of the partners, so that state officials, via scrolling, seek to allocate orphans to these couples. Similarly, Figure \ref{fig:exemplo-sistemas} B illustrates another example of a system that can skew a person’s decision making. In the Brazilian electoral period, the superior electoral court created a system that shows the alphabetical list of all candidates for the election \cite{brazilian-system}. This practice is also reproduced by the main news portals of the country (e.g. www.globo.com, www.uol.com, www.estadao.com, etc.) for helping undecided voters to look for a candidate. 

Despite this context, we are not aware of any research work that investigated whether the simple act of displaying people's names in alphabetical order is capable of causing imbalances with respect to gender, which might induce biased choices. By imbalances we mean that for a particular screen page in which part of a list of persons is displayed, the proportion of men and women in this page is different from the proportion of the entire list. We aim at alerting software engineers and HCI experts about the need of specific information visualization guidelines that avoid to show unbalanced lists and in which circumstances this may provoke prejudice. This goal motivated us to answer the following research questions:

\begin{itemize}
\item {\verb|RQ1|}: For a list of people ordered by first names, is there any imbalance in the first pages that might favor or penalize some gender? 
\item {\verb|RQ2|}: Does  gender imbalance vary as a function of the origin of the names?
\item {\verb|RQ3|}: Does gender imbalance depend on features of the list to be displayed (e.g. size of the list, the proportion of men and women of the entire list) ?
\item {\verb|RQ4|}: Does gender imbalance depend on the size of the screen device (e.g. computer, smartphone, tablet), which determines the size of the list that is displayed in the  pages?

\end{itemize}

The empirical approach to analyze the research questions relies on real-world  data analysis.  In order to answer RQ1 and RQ2 we explore the characteristics of datasets (named here primary datasets) of first names of the population of five countries. Each dataset has the frequency of adoption of the first name in the country. We use the statistical parity technique and a fairness metric to identify from samples of individuals from each country’s datasets, the imbalances that can induce bias [26]. 

 We quantify the risk that a sample of $N$ persons, when placed alphabetically by first name, shows an imbalance in the {\it Top-k} ($k \leq N$ represents the amount of persons who are displayed in a screen page) individuals of the population in relation to gender. In other words, this metric increases its value as function of the low representativeness of a gender, in the {\it Top-k} individuals. For answering RQ3 and RQ4, the calculation of the fairness metric relies on the variation of samples of the primary datasets containing the first names, organized by origin (country). The variation is estimated by  the size of the samples as well as the proportion of men and women. Doing so, it was possible to highlight in which circumstances there is a greater risk of bias. 
 
 The main findings of this research are the following:

\begin{itemize}
\item {\verb|Finding 1|}: Displaying the alphabetical ordering of names of people may imply  substantial gender imbalances.When {\it k}, that represents the size of the first page of a device, is very small, (e.g.,  $k \leq 15$),  it is observed that there are statistically fewer individuals of the female gender for four different countries(except Brazil), demonstrating a low representation of women among the first elements of these samples. This observation regarding the value of $k$  is particularly relevant for HCI,  given the typical size of  device screens.  

\item {\verb|Finding 2|}: Regarding the origin of the first names, we observed that gender imbalances are more likely to occur in samples from datasets of individuals from Brazil and Spain.

\item {\verb|Finding 3|}: We show that samples with fractions of men and women similar to the fractions of the primary dataset are more likely to have a gender imbalance. We show that the larger the sample, the greater the imbalance. This occurs because of the occurrence of large sequences of repeated names of same gender that ends up to dominate certain screen pages. When this occurs at the first pages, gender is imbalanced and may lead to biased decisions

\item {\verb|Finding 4|}: As a consequence of finding \#1, gender imbalance occurs in all devices for all primary datasets, because small values of $k$ have high imbalances).
\end{itemize}

The results emphasize the importance of guidelines on information visualization to be followed by software engineers, HCI designers and practitioners in order to avoid unbiased choices with respect to gender from the display of alphabetically ordered names of people,  We are in line with HCI works and standard [ISO 9241-210] that claim a broader view on usability beyond efficiency aspects.  One must also capture features as users' perception \cite{hassenzahl2007hedonic, hassenzahl2015experience}, emotion \cite{hassenzahl2018thing}, privacy \cite{culnan2000protecting}, and cultural issues of a modern and sophisticated society, such as ethics.

\section{MOTIVATION: THE CASE OF LISTING POLITICAL CANDIDATES
}
\begin{figure*}[t]
\centering
\includegraphics[width=1\linewidth]{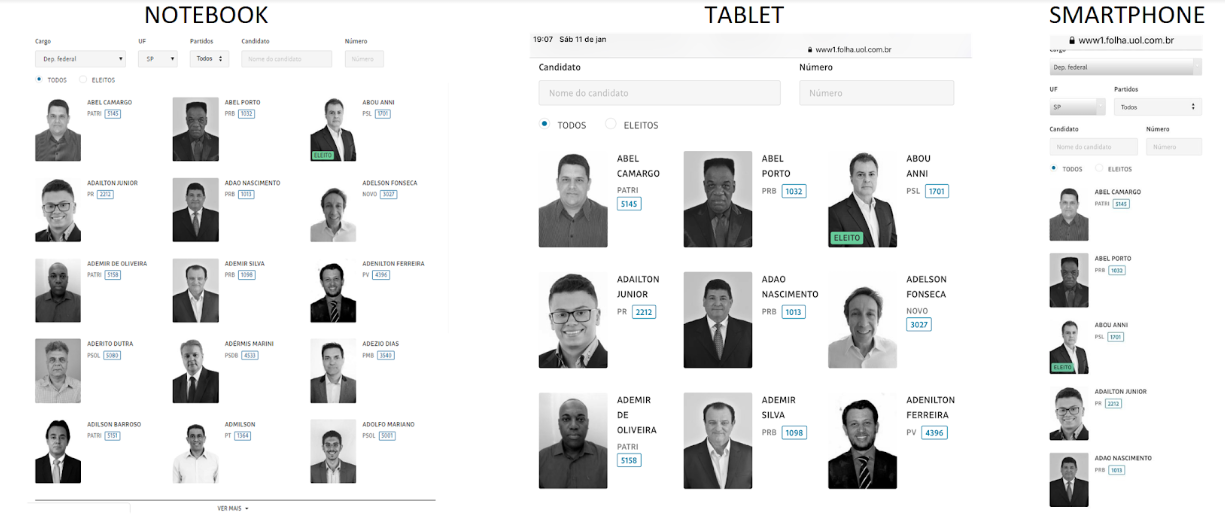}
\caption{
Figure 2. Examples of the first page after consulting the list of candidates from the State of São Paulo for the Brazilian Parliament from three different devices. The values of $k_1$ are 5, 9 and 15.}
\label{fig:candidatos-devices}
\end{figure*}

In this section, we present a real case study that shows the relevance of the problem discussed here.  During the electoral periods,  Brazilian News  portals (e.g.,  Globo Network, UOL,  etc.) and official sites, such as the one of the Supreme Election Court, list the names of politicians running for office.  Usually,  the sites  list candidates in alphabetical order.  Our case study focuses on candidates for parliament. Brazilian electoral legislation requires that each political party have at least 30\% female candidates for each state of the Federation. As the choice of representatives is made by state, all sites allow the consultation of candidates by state or political party. Even when a user applies a filter by party or location, the list of candidates is shown in alphabetical order. If a user is using a computer screen to view candidates, the site typically returns the first fifteen candidates (see example in Figure \ref{fig:exemplo-sistemas}). If the query is being made from a smartphone, the number of candidates  shown is smaller, about 5 per screen. Figure \ref{fig:candidatos-devices} shows an example of the same query made on the Folha Portal1 from a notebook, tablet and  smartphone. Due to the variability of the size of the devices, we decided to conduct a study with the first page, k1, assuming the values of 5 for smartphone, 9 for tablet and 15 for notebook.  We investigate gender unbalanced in the list of candidates in the first page of these different devices.

Table \ref{tab:table_candidates} shows the results of the percentage of women on the front page for each of the 26 States of the Brazilian Federation for the Federal Parliament (left) and State Parliament (right). The column size indicates the number of candidates of each State. The lists of candidates for the State Parliament are always greater than the candidate list for the federal parliament. The number of women and men in the first page was expected to be equal to the proportion of the list of candidates (greater than or equal to 30\% of women as required by law). The column $Perc_{fd}$ indicates the percentage of women candidates for the State. The results  of Table \ref{tab:table_candidates} show no evidence that the balanced ratio required by the legislation was achieved. Quite the contrary, the lists of the first pages for all States with $k_1$ = $5$, $9$ and $15$ exhibit a different proportion of gender. Regarding candidates for the Federal Parliament, seven States have the percentage of women lower than expected, for all values of $k$ (AL, AM, BA, PR, RJ, SP,TO, in red at Table \ref{tab:table_candidates}). In the two most populous states (São Paulo-SP, Rio de Janeiro-RJ) with long lists of candidates, the percentage of women is lower, being zero in SP for all values of k. For the State Parliaments, the gender imbalance is even bigger. In $14$ States (AC, AP, BA, ES, GO, PA, PE, PR, RJ, RO, SE, SP, TO) the percentage of women is lower than expected for all devices.  The number of times in which the percentage of women is lower than expected is 38 for the Federal Parliament and 55 for the State Parliaments. The  results indicate  the size of the list of candidates might be a factor that influences the unbalance, as we discuss later in the paper.

\setlength{\tabcolsep}{2pt}
\begin{table}
\centering
\caption{\label{tab:table_candidates}Women percentage ($Perc_{f}$)  in the of candidates for the Federal Parliament and State Parliaments who are displayed in the first page for different values of $k_1$, that represent different  devices, i.e., smartphone, tablet and notebook.}

\begin{tabular}{ll}

\begin{tabular}{cccccc}
\multicolumn{6}{c}{Federal Parliament}\\
\toprule

State & $k_{1}$ = 5 & $k_{1}$ = 9 & $k_{1}$ = 15 & $size$ & $Perc_f$       \\
\bottomrule

AC & \cellcolor{blue!25}0.4 & \cellcolor{red!25}0.33 & \cellcolor{red!25}0.33 & 83 & 0.34\\
AL & \cellcolor{red!25}0.2 & \cellcolor{red!25}0.11 & \cellcolor{red!25}0.2 & 81 & 0.33\\
AM & \cellcolor{red!25}0.0 & \cellcolor{red!25}0.22 & \cellcolor{red!25}0.27 & 142 & 0.31\\
AP & \cellcolor{blue!25}0.4 & \cellcolor{blue!25}0.33 & \cellcolor{red!25}0.27 & 112 & 0.33\\
BA & \cellcolor{red!25}0.2 & \cellcolor{red!25}0.22 & \cellcolor{red!25}0.2 & 481 & 0.32\\
CE & \cellcolor{blue!25}0.4 & \cellcolor{blue!25}0.33 & \cellcolor{red!25}0.27 & 253 & 0.32\\
DF & \cellcolor{red!25}0.2 & \cellcolor{blue!25}0.44 & \cellcolor{blue!25}0.47 & 183 & 0.32\\
ES & \cellcolor{blue!25}0.6 & \cellcolor{blue!25}0.44 & \cellcolor{red!25}0.33 & 164 & 0.34\\
GO & \cellcolor{blue!25}0.4 & \cellcolor{blue!25}0.44 & \cellcolor{blue!25}0.47 & 212 & 0.33\\
MA & \cellcolor{blue!25}0.4 & \cellcolor{blue!25}0.56 & \cellcolor{blue!25}0.53 & 202 & 0.34\\
MG & \cellcolor{red!25}0.2 & \cellcolor{blue!25}0.33 & \cellcolor{blue!25}0.6 & 889 & 0.31\\
MS & \cellcolor{red!25}0.2 & \cellcolor{blue!25}0.33 & \cellcolor{blue!25}0.4 & 122 & 0.32\\
MT & \cellcolor{red!25}0.2 & \cellcolor{blue!25}0.56 & \cellcolor{blue!25}0.4 & 136 & 0.35\\
PA & \cellcolor{blue!25}0.4 & \cellcolor{blue!25}0.44 & \cellcolor{blue!25}0.4 & 132 & 0.3\\
PB & \cellcolor{red!25}0.2 & \cellcolor{blue!25}0.33 & \cellcolor{red!25}0.27 & 149 & 0.31\\
PE & \cellcolor{red!25}0.2 & \cellcolor{blue!25}0.33 & \cellcolor{blue!25}0.4 & 347 & 0.32\\
PI & \cellcolor{blue!25}0.6 & \cellcolor{blue!25}0.33 & \cellcolor{red!25}0.27 & 134 & 0.33\\
PR & \cellcolor{red!25}0.0 & \cellcolor{red!25}0.11 & \cellcolor{red!25}0.13 & 434 & 0.3\\
RJ & \cellcolor{red!25}0.2 & \cellcolor{red!25}0.11 & \cellcolor{red!25}0.07 & 1119 & 0.31\\
RN & \cellcolor{blue!25}0.4 & \cellcolor{blue!25}0.44 & \cellcolor{red!25}0.27 & 119 & 0.34\\
RO & \cellcolor{blue!25}0.4 & \cellcolor{red!25}0.22 & \cellcolor{blue!25}0.4 & 106 & 0.34\\
RR & \cellcolor{blue!25}0.6 & \cellcolor{blue!25}0.44 & \cellcolor{blue!25}0.53 & 144 & 0.33\\
RS & \cellcolor{blue!25}0.6 & \cellcolor{blue!25}0.44 & \cellcolor{blue!25}0.33 & 414 & 0.33\\
SC & \cellcolor{blue!25}0.4 & \cellcolor{red!25}0.22 & \cellcolor{blue!25}0.33 & 240 & 0.32\\
SE & \cellcolor{blue!25}0.4 & \cellcolor{blue!25}0.33 & \cellcolor{red!25}0.27 & 117 & 0.32\\
SP & \cellcolor{red!25}0.0 & \cellcolor{red!25}0.0 & \cellcolor{red!25}0.0 & 1603 & 0.31\\
TO & \cellcolor{red!25}0.2 & \cellcolor{red!25}0.11 & \cellcolor{red!25}0.2 & 82 & 0.34\\

\bottomrule

\hline
\end{tabular}

&

\begin{tabular}{cccccc}
\multicolumn{6}{c}{State Parliament}\\
\toprule

State & $k_{1}$ = 5 & $k_{1}$ = 9 & $k_{1}$ = 15 & $size$ & $Perc_f$       \\
\bottomrule

AC & \cellcolor{red!25}0.0 & \cellcolor{red!25}0.0 & \cellcolor{red!25}0.27 & 445 & 0.32\\
AL & \cellcolor{red!25}0.2 & \cellcolor{red!25}0.22 & \cellcolor{blue!25}0.47 & 321 & 0.32\\
AM & \cellcolor{blue!25}0.4 & \cellcolor{blue!25}0.33 & \cellcolor{red!25}0.27 & 613 & 0.3\\
AP & \cellcolor{red!25}0.0 & \cellcolor{red!25}0.0 & \cellcolor{red!25}0.2 & 462 & 0.33\\
BA & \cellcolor{red!25}0.0 & \cellcolor{red!25}0.11 & \cellcolor{red!25}0.07 & 610 & 0.3\\
CE & \cellcolor{red!25}0.0 & \cellcolor{red!25}0.22 & \cellcolor{blue!25}0.33 & 578 & 0.3\\
DF & \cellcolor{blue!25}0.4 & \cellcolor{blue!25}0.44 & \cellcolor{red!25}0.27 & 949 & 0.31\\
ES & \cellcolor{red!25}0.2 & \cellcolor{red!25}0.22 & \cellcolor{red!25}0.2 & 590 & 0.31\\
GO & \cellcolor{red!25}0.2 & \cellcolor{red!25}0.11 & \cellcolor{red!25}0.2 & 828 & 0.31\\
MA & \cellcolor{blue!25}0.6 & \cellcolor{blue!25}0.33 & \cellcolor{red!25}0.27 & 496 & 0.32\\
MG & \cellcolor{red!25}0.2 & \cellcolor{red!25}0.11 & \cellcolor{blue!25}0.33 & 1329 & 0.32\\
MS & \cellcolor{red!25}0.2 & \cellcolor{blue!25}0.33 & \cellcolor{blue!25}0.33 & 337 & 0.33\\
MT & \cellcolor{red!25}0.0 & \cellcolor{red!25}0.0 & \cellcolor{red!25}0.07 & 338 & 0.31\\
PA & \cellcolor{red!25}0.2 & \cellcolor{red!25}0.22 & \cellcolor{red!25}0.13 & 656 & 0.32\\
PB & \cellcolor{blue!25}0.6 & \cellcolor{blue!25}0.44 & \cellcolor{blue!25}0.6 & 400 & 0.31\\
PE & \cellcolor{red!25}0.2 & \cellcolor{red!25}0.22 & \cellcolor{red!25}0.2 & 644 & 0.31\\
PI & \cellcolor{blue!25}0.4 & \cellcolor{blue!25}0.33 & \cellcolor{blue!25}0.4 & 211 & 0.32\\
PR & \cellcolor{red!25}0.0 & \cellcolor{red!25}0.0 & \cellcolor{red!25}0.07 & 751 & 0.31\\
RJ & \cellcolor{red!25}0.2 & \cellcolor{red!25}0.22 & \cellcolor{red!25}0.27 & 2368 & 0.31\\
RN & \cellcolor{blue!25}0.4 & \cellcolor{blue!25}0.33 & \cellcolor{blue!25}0.53 & 315 & 0.33\\
RO & \cellcolor{red!25}0.0 & \cellcolor{red!25}0.0 & \cellcolor{red!25}0.0 & 408 & 0.32\\
RR & \cellcolor{red!25}0.2 & \cellcolor{blue!25}0.33 & \cellcolor{blue!25}0.47 & 441 & 0.31\\
RS & \cellcolor{blue!25}0.4 & \cellcolor{red!25}0.22 & \cellcolor{blue!25}0.4 & 829 & 0.32\\
SC & \cellcolor{blue!25}0.4 & \cellcolor{blue!25}0.33 & \cellcolor{red!25}0.27 & 446 & 0.32\\
SE & \cellcolor{red!25}0.2 & \cellcolor{red!25}0.22 & \cellcolor{red!25}0.2 & 301 & 0.33\\
SP & \cellcolor{red!25}0.2 & \cellcolor{red!25}0.22 & \cellcolor{red!25}0.2 & 2055 & 0.31\\
TO & \cellcolor{red!25}0.2 & \cellcolor{red!25}0.22 & \cellcolor{red!25}0.2 & 217 & 0.31\\

\bottomrule

\hline
\end{tabular}
\end{tabular}
\end{table}

\section{Primary Datasets}

In order to generalize the study previously presented, we  chose five  primary datasets that include the first names of people from that were born  or reside in five countries. In the five datasets, there is an association of the first name with the number of individuals  with that name and the gender  associated with it. The datasets include a large number of names from the French, American, Spanish, Brazilian and Scottish populations.

The dataset from France comes from the INSEE (National Institute of Statistics and Economic Studies) which collects, produces, analyzes and disseminates information on the French economy and society. The INSEE  provides an interactive tool \cite{data-franca} that allows for access for each year, the ranking of the 100 most commonly registered first names for female and for male babies. Therefore, in order to represent the French population, whose life expectancy is 82 years, according to INSEE, we look at  rankings from 1935 to 2017, with 47,896,230 people (71\% of the total population), 25,247,187 ($\approx 53\%$) men and 22,457,315 ($\approx47\%$) women.

For the United States, the primary dataset comes from  the website of the Social Security Administration \cite{data-usa}. On the SSA platform, there is one file per year with the first names and gender of the American babies. For privacy reasons, first names with fewer than five occurrences are not included in the dataset. For this study, we used the files with records from births from 1939 to 2017, since the life expectancy of the average American is 78 years. Thus, considering the US population of 325,147,121 in 2017 \cite{info-pop-usa},  about 85\% (279,034,205) of the population is covered, being 143,895,912 (52\%) men and 135,138,293 (48\%) women.

In Scotland, there is a government department, the National Records of Scotland \cite{data-escocia}, that provides records from births for men and women since 1974. There are 2,727,214 million individuals, representing about $\approx50\%$ of the total population. In the primary dataset there are 1,399,018 men ($\approx51\%$) and 1,328,196 ($\approx49\%$) women.

Data from Spain comes from the National Statistics Institute \cite{data-esp}. We have accessed a spreadsheet that has all the first names with a frequency greater than or equal to 20 of the residents of Spain. It covers 43,481,592 people (93\%) of the total population, being 21,471,601 men and 22,009,991 women.

In the case of Brazil, the data were obtained from the Brazilian Institute of Geography and Statistics (IBGE), in the project "Names in Brazil", that makes available the number of people with each name and their respective gender. In the dataset of Brazil, like we did with France and the United States, the life expectancy (75 years) was used to represent the current living population, so that data from the 1940s to 2000s were considered for the country and has 76,098,733 people, being 42,986,632 men (57\%) and 33,112,101 women (43\%).

\section{Methodology}

Using the primary datasets previously described, we examined the following research question:  given an alphabetically ordered sample of $N$ individuals, randomly taken from a primary dataset of individuals, can we quantify  the gender imbalance in  the list  of the  {\it Top-k} individuals ($k \leq n$). 
The methodology consists of exploring samples of individuals in different proportions of women and men. The intuition behind this strategy is that the samples represent the numerous files with first names of people that can be used by information systems such as  the list of candidates to the Brazilian Parliament , that represents a sample of names. We created two groups of samples. The first one tended to maintain the gender ratio of the primary dataset, while in the second group of unbalanced samples were generated in order to emulate contexts where there is  a historical prevalence of one gender over another (e.g., military enlistment).The samples were generated using the Fisher-Yates shuffle technique \cite{saeed2014fisher} to generate a random permutation of a finite linear array.This algorithm was chosen because it guarantees an unbiased result that is, every permutation of the array is equally likely. For each primary dataset, the algorithm receives as parameter:the set of individuals of the primary dataset I; The percentage of females of the sample $Perc_{fs}$; and the size of the sample n(number of individuals). Initially, 100 samples of 1000 individuals ($n=1000$) were generated for values of $Perc_{fs}$ belonging to {0.05, 0.1, 0.15, 0.2, 0.25,0.3, 0.35, 0.4, 0.45, 0.5, 0.55, 0.6, 0.65, 0.7, 0.75, 0.8, 0.85, 0.9, 0.95}. Note that samples in which $Perc_{fs}$ has the value 0.3 simulates the Brazilian political scenario of the candidates where 30\% of the individuals are women.

 We analyzed with statistical parity techniques \cite{yang2017measuring} the two groups of samples to verify if the alphabetical ordering of individuals’ first names causes gender imbalances in the {\it Top-k} individuals. This  technique consists of verifying whether the sample has proportion of males and females similar to the demographics of the entire population described in the primary datasets.
 
In addition to the statistical analysis, we adapted a ranking fairness metric used in \cite{yang2017measuring} to our problem. This metric is the normalized discounted difference ({\it rND}),  whose value increases when there are disproportions in the studied trait (i.e.,  gender) among the {\it Top-k} elements (i.e., names) of the sample. Let $S^f$ represent the set of women while the set of men is represented by $S^m$. The metric makes a logarithmic discount  based on the position $k$ of the elements in the sample. In addition, it quantifies the relative representation of $S^f$ in a ranking, in this way, the equity will be calculated based on discrete points of the ranking ({\it Top-10}, {\it Top-20}, etc.).

Normalized discounted difference ({\it rND}) calculates, for each $k$, the difference from the proportion of women for the first $k$ individuals of the sample to the same proportion for the complete sample. Formally, {\it rND} can be defined as:

\begin{equation}
rND = \frac{1}{Z} \sum_{{k=10,20,...}}^{N} \frac{1}{\log_2 k} \left|\frac{size(S_{k}^f)}{k} - \frac{size(S^f)}{N}\right|
\end{equation}

Where {\it size($S_{k}^f$)} represents the amount of women in the $k$ first individuals of the sample and { \it size($S^f$)} represents the total amount of women in the sample. The values are accumulated in discrete points in the ranking representing the different values of $k$. The value of $k$ starts in $10$ and varies in intervals of $10$. A logarithmic discount is applied and the value is normalized. Normalizer $Z$ is defined as the highest value of {\it rND} found. The {\it rND}, presented in this Section, is normalized for values between [$0$, $1$]. It has its best (fairest) value at $0$ and its worst value at $1$.

\section{Measuring Imbalance Via Data Analysis}

\begin{figure*}[t]
\centering
\includegraphics[width=0.72\linewidth]{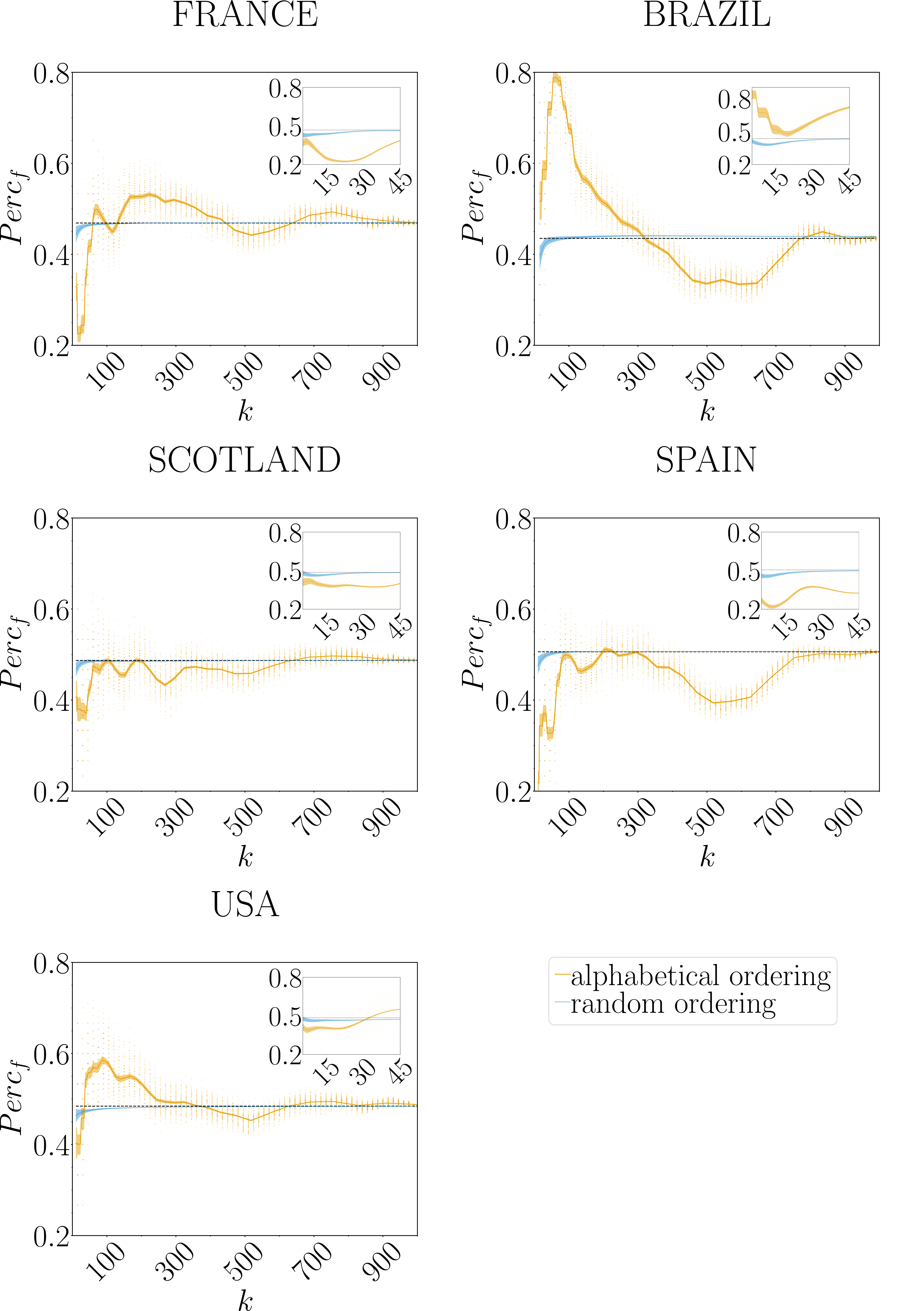}
\caption{
Percentage of women among the {\it Top-k} individuals in the samples.
}
\label{fig:experimento1}
\end{figure*}

 In order to answer for research question \#1 (RQ1), we look at  Figure \ref{fig:experimento1}, that exhibits the percentage of women ($Perc_{f}$) among the {\it Top-k} individuals in the $100$ samples of $1000$ people ($n=1000$), randomly extracted  from each primary dataset. There are two curves.  The blue curve represents the samples of individuals that were randomly ordered while the orange one shows  individuals that were  alphabetically ordered. The orange solid line illustrates the result of a non parametric regression (using Nadaraya-Watson method) \cite{nadaraya1964, watson1964}) over the orange dots ($Perc_{f}$ values in the samples ordered alphabetically). The solid blue line illustrates the result of a regression, also using the Nadaraya-Watson method, for the $Perc_{f}$ values when the sample is randomly ordered (the dots of the random sample were omitted in the Figure for sake of clarity). In the  experiments with the five primary datasets, it was observed that a random ordering maintains the proportion of  women  with respect to the primary dataset (i.e., dashed black line), among the first $k$ individuals. In contrast, alphabetical ordering in the sample datasets was shown to generate a gender imbalance in the population. The most striking cases occurred in France ($k=23$), Scotland ($k=37$), Spain ($k=10$), Brazil ($k=59$) and USA ($k=86$). For the three first datasets the striking cases represent women outnumbering men. The other two represent the contrary. For $k<15$ it is noted that alphabetical order tends to significatively put more men than women in the list (exceeding the limits of the 95\% confidence intervals, estimated by the bootstrap method \cite{racine2004}). For $k < 50$ this also occurs except for the Brazilian dataset. 
 
The unbalancing effect is because samples reflect two features of the Primary Datasets. First, in all datasets, for any letter of the alphabet, it may occur popular names adopted by many people. Ordering a sample means defining what popular name comes first as well as generating an ordered alternance between popular names. When in the {\it Top-k} of the sample there is a popular name, the balancing tends do be dominated by the gender of the popular name. For example, if there is a popular male name starting with “A” that comes first than any female popular name, men tend to outnumber women for small values of $k$, because the {\it Top-k} will be dominated by men with this name. 

Besides this effect of popularity, it may occur consecutive different names of either men or women what also implies to large sequences of names of same gender. These two features combined explain the alternance of dominance leading to a non-monotonic effect that can be seen in all Orange curves of the graphs of Figure \ref{fig:experimento1} (except for Spain in which male names dominate for all values of $k$).

%All graphs show on the y-axis the percentage of women ($Perc_{f}$) in the $k$($x-axis$) first elements of the sample in an experiment where 100 random samples of size n=1000. Orange dots correspond to the values of $Perc_{f}$ when the samples are sorted alphabetically. The orange solid line illustrates the result of a non parametric regression (using Nadaraya-Watson method) \cite{nadaraya1964, watson1964}) over the orange dots and the orange shadow represents the 95\% confidence interval of this regression estimated by the bootstrap method \cite{racine2004}. The solid blue line illustrates the result of a regression, also using the Nadaraya-Watson method, for the $Perc_{f}$ values when the sample is randomly ordered. The blue shadow represents the 95\% confidence interval of the regression made on the $Perc_{f}$ values when the sample is randomly ordered, this confidence interval was estimated using the bootstrap method. The black dotted line indicates the percentage of women considering the data from the primary dataset of the country.

\begin{figure}[t]
\centering
\includegraphics[width=0.8\linewidth]{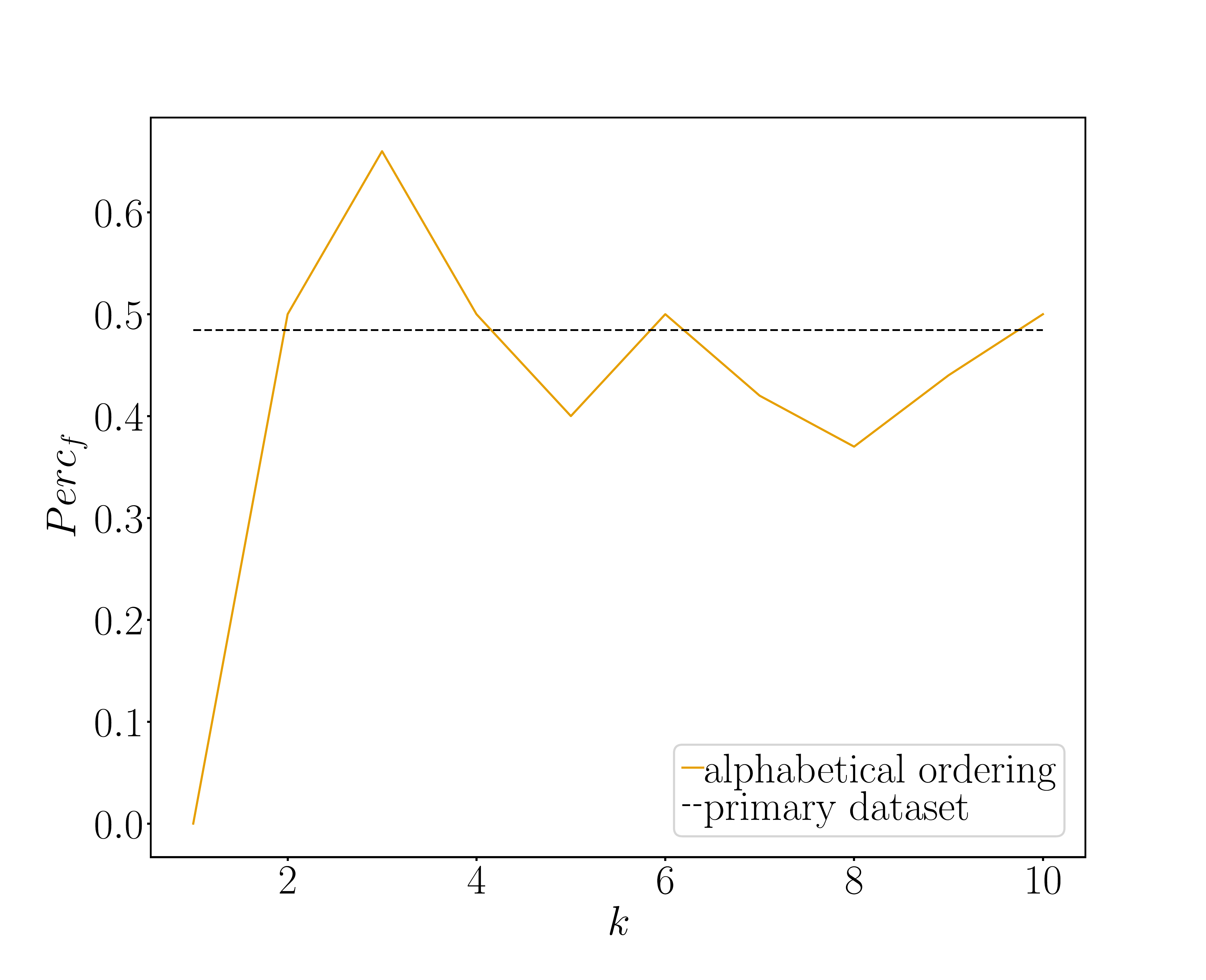}
\caption{$Perc_{f}$ behavior for the sample illustrated in Table \ref{tab:exemplo-experimento} B(sample with intercalated gender).}
\label{fig:exemplo-intercalacao}
\end{figure}

The variation of the orange curves occurs because alphabetical ordering promotes the emergence of sequences from individuals with the same name that end up interspersed with sequences of individuals of another name and different gender. Suppose the sample containing only $10$ individuals illustrated from Table \ref{tab:exemplo-experimento}, there are three people named Brian. In column A, names appear randomly ordered.   The three ``Brians’  are in positions that are distant from each other, i.e., one in the first position, the other one in fourth and the third one in the tenth position. On the other hand, when the names are alphabetically ordered (column B), they are in contiguous positions, which significantly decreases  the $Perc_{f}$ in the sample. Therefore, as can be seen there are other sequences of the same name  in the sample. For example, sequences of "Amy" and "Christina"  cause  changes to the $Perc_{f}$ values,which produces the non-monotonic behavior also shown in Figure \ref{fig:exemplo-intercalacao}. 

\begin{table}
\centering
  \caption{Sample to explain the non-monotonicity behaviour of $Perc_{f}$.The samples have the same individuals, however, in A they are randomly ordered while in B they are alphabetically ordered}
  \label{tab:exemplo-experimento}
  \begin{tabular}{p{0.5cm}p{1.6cm}p{1.2cm}p{0.9cm}|p{0.5cm}p{1.6cm}p{1.2cm}p{0.7cm}}
    \multicolumn{4}{l}{A} & 
    \multicolumn{4}{l}{B} \\
    \hline
    $k$&Name&Gender&$Perc_{f}$&$k$&Name&Gender&$Perc_{f}$\\
    \hline

1&Brian&Male&0.00&1&Aaron&Male&0.00\\
2&Aaron&Male&0.00&2&Amy&Female&0.50\\
3&Christina&Female&0.33&3&Amy&Female&0.66\\
4&Brian&Male&0.25&4&Andrew&Male&0.50\\
5&Christina&Female&0.40&5&Ashley&Female&0.50\\
6&Amy&Female&0.50&6&Brian&Male&0.40\\
7&Ashley&Female&0.57&7&Brian&Male&0.42\\
8&Amy&Female&0.62&8&Brian&Male&0.37\\
9&Andrew&Male&0.55&9&Christina&Female&0.44\\
10&Brian&Male&0.50&10&Christina&Female&0.50\\
\hline
\end{tabular}
\end{table}

\begin{figure}[t]
\centering
\includegraphics[width=1\linewidth]{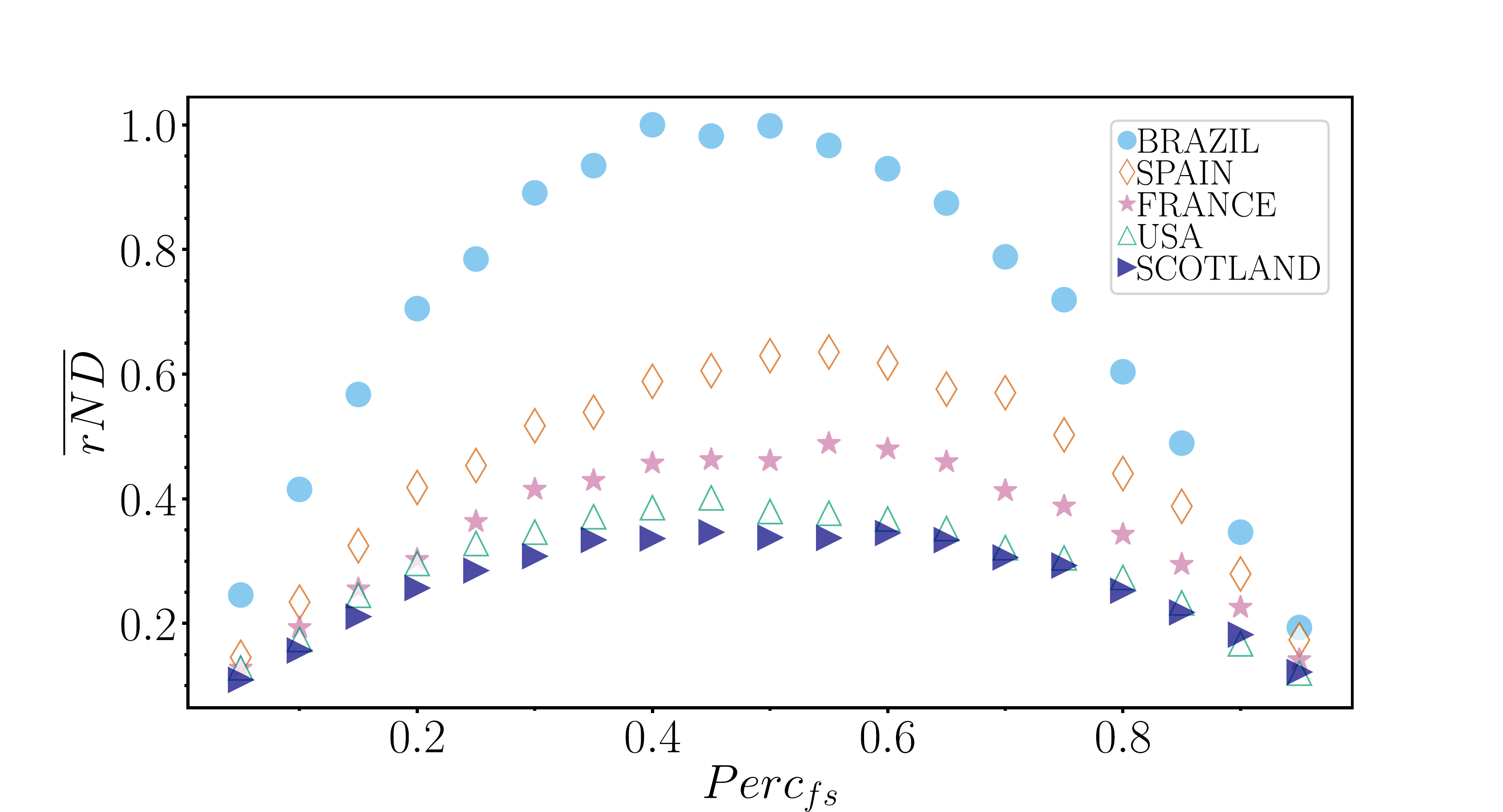}
\caption{Relationship between the {\it rND} and the percentage of women in the samples. The blue circles, orange diamonds, pink stars, green triangles and purple right triangles represents, respectively, the values of {\it rND} for Brazil, Spain, France, United States and Scotland. When the proportion between men and women tends to be equal ($Perc_{fs} \approx 0.5$) the values of {\it rND} are higher.
}
\label{fig:rnd-percf}
\end{figure}

Figure \ref{fig:rnd-percf} helps to answer RQ2. It shows that in samples generated from the Brazilian primary dataset, the values of {\it rND}(blue circles) are higher than those computed to other languages. This result reveals that, among the datasets studied, there is a greater risk of gender bias in information systems that manipulate first names of Brazilian and Spanish origin than systems that use names from other countries. This is especially so because of the low representation of a gender (male in Brazil and female in Spain), especially in the first individuals of the samples. It is worth noting that Latin America (Portuguese and Spanish speakers) is the region where listing persons in alphabetical ordering of first name is widely used. 

Figure \ref{fig:rnd-percf} helps also to answer RQ3 because it shows the effect that alphabetical ordering has on the disproportion of a given gender among the first $k$ individuals of unbalanced samples (samples with quite different gender distribution than its primary dataset). We have generated samples with 5\%, 10\%, 15\%, ..., 85\%, 90\% and 95\% proportion of women in relation to men. For each percentage, $100$ samples of $1000$ individuals($n=1000$) were generated, these were alphabetically ordered and their {\it rND} mean value was calculated. Finally, the mean values of {\it rND} were normalized to $Z=1.1$, the highest {\it rND} value found.

The  relation between the fairness metric ({\it rND}) and the percentage of women in the samples ($Perc_{fs}$) is plotted in Figure \ref{fig:rnd-percf}. It can be observed that there is a tendency to assign higher values of {\it rND} to samples in which the proportion of men and women are similar. When the percentage of women ($Perc_{fs}$) is 50\% the values of {\it rND} are the greatest.  This result reveals that samples having similar proportion of men and women, after being alphabetically ordered, tend to favor one gender over the other in {\it Top-k} individuals

\begin{figure}[t]
\centering
\includegraphics[width=0.8\linewidth]{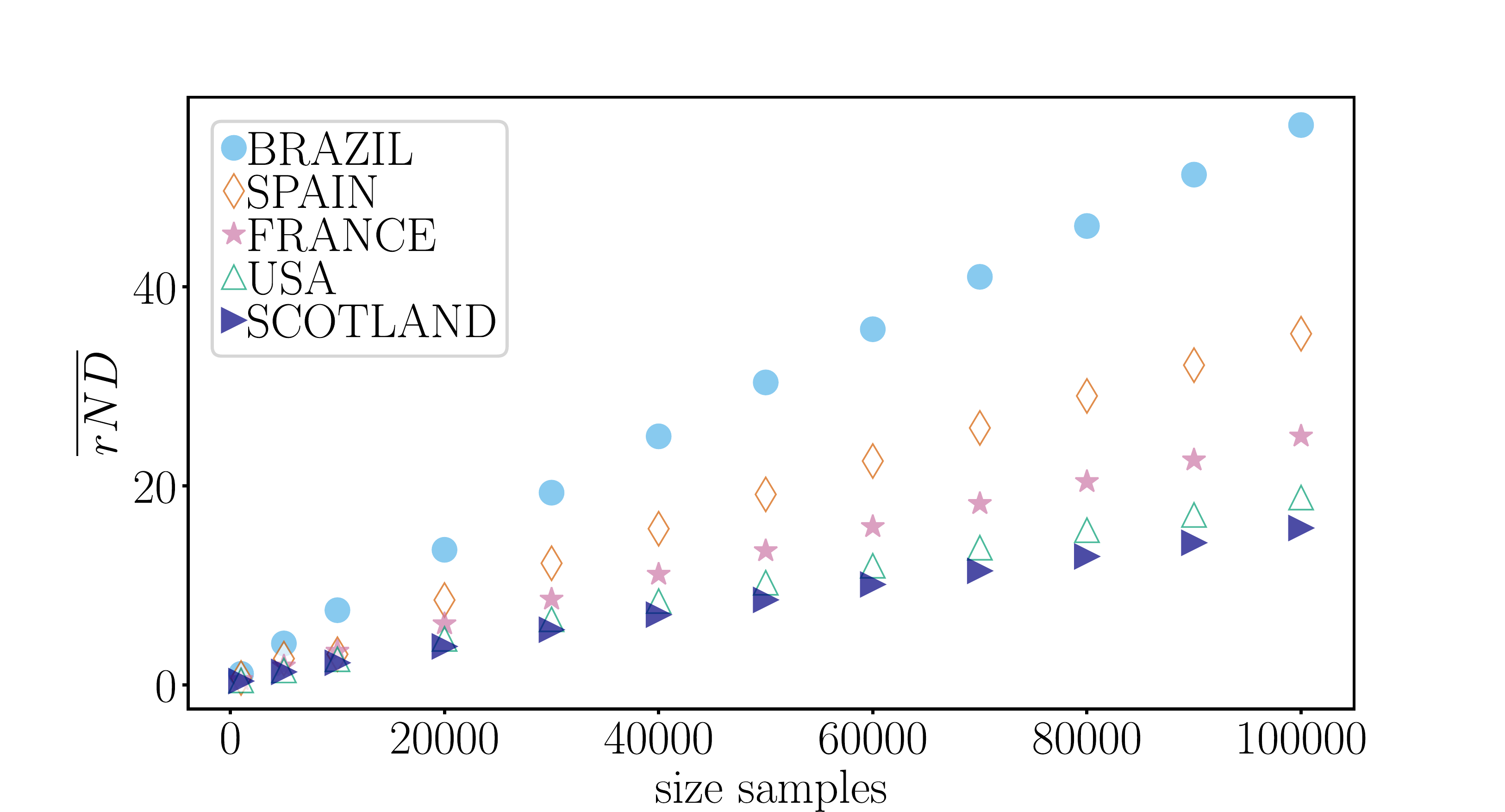}
\caption{ Average {\it rND} from 100 samples of each primary dataset for different values of the sample size $n$. The graphs illustrate in the {\it y-axis} the average {\it rND} and at the {\it x-axis} the values of n. In the five countries it was observed that the greater the n, greater the {\it rND}. }
\label{fig:rnd-size}
\end{figure}

Another finding related to research question \#3  is illustrated in Figure \ref{fig:rnd-size}, that shows the average {\it rND}  as function of  the size of the samples.   The average {\it rND} is computed for 100 samples of each primary dataset. We observe that the greater the sample, the greater the average {\it rND}.  The size of the sample (n=1000) used in the experiment and depicted in Figure \ref{fig:experimento1} was a conservative choice that is not too biased. The results shown in Figure \ref{fig:rnd-size} are explained by the fact that the greater the dataset the greater the number of repeated names.  As a  consequence,  we perceive large sequences of names of same gender that end up to dominate certain screen pages. When this occurs at the first pages, choices may be biased for that gender represented in these pages.

Finally, the results depicted in Figure \ref{fig:experimento1} allows to infer that the risk of disproportion of men and women is independent on device (RQ4). This is due to the fact that this occurs at $k$ values lesser than $17$ (see zoom in each country chart), which would include the first page of smartphones, tablets and even personal computers.

\section{Implications On Information Visualization}

Experts on usability have been studying  methods for effective user navigation in screens \cite{nielsen1999designing} and the role of the order to present elements of information  on the screen. Basically  they  have assessed  the preference of  people to sites that go straight to the point, letting them get things done quickly. \cite{pemberton2003hotel} suggests that improved usability is about optimizing the time that  users take to achieve their purpose, how well they achieve it, and the satisfaction in doing it. Portals as those we have exemplified in this article are typical examples of Web listing interfaces in which the scrolling task is done from the users click in a specific area of interaction. For them the users’ attention is more dedicated to the first page (usually called above the fold) than in the rest. In recent usability tests, using new techniques, such as infinite scrolling, eye tracking data showed that, people scroll vertically more than they are used to but they will still look more above the page fold than below it. They spend about 80\% of time in the first three pages. That is the reason we have prioritized the impact of the first pages.

 Two basic theories explain the behavior of the users in the previously cited usability tests. The Fitts's Law \cite{fitts1954information, mackenzie1992fitts} predicts that the time required to rapidly move to a target area is a function of the ratio between the distance to the target and the width of the target. On the other hand, the information foraging theory (IFT) proposed by \cite{pirolli2007information} helps to understand how humans search for information on the Web. Both theories address the notion of cost of searching information in contrast with the benefits or utility of having that information. In controlled experiments with human users browsing Web pages for a particular task, IFT shows that the frequency of access to Web pages follows a logarithmic distribution. The first pages represent 80\% of the access while the rest of the pages represents the other 20\%. The reason behind this is that the expected values of continuing to scan diminishes with each additional listing scanned. If the list of results is very long, there is usually a point at which the information seeker (the forager in the IFT vocabulary) faces the decision of whether it is worth the effort of continuing to search for a better result than anything encountered so far. 
 
For the example of access to the political candidates portal, discussed in the second Section of this paper,  the information foraging theory  models the user behavior as follows. There are basically two types of tasks that voters who access portals with the candidate list can perform. The first task is to consult the number that identifies the candidate. In Brazil  voters have to enter the identification of the candidate in the electronic ballot box. In this case the alphabetical order has no impact, because the candidate has already been chosen by the voter. A second task is that one performed by voters who have no candidate yet. In this case, access to electoral portals is exploratory in order to seek information about the candidates. Portals basically present  the candidate's name, political party, number and photo. The voter may decide to cast her vote solely based on this information or browse other sites for more information to support her choice, such as the candidate's previous performance on the job. Both scenarios are typically described as information foraging processes and consequently the names that appear on the front pages tend to be privileged. If there is a gender imbalance between the names that appear first, biased decisions could occur.

 Listing people alphabetically on a computer screen or on any other electronic output device is a traditional and universal technique. Alphabetical order is easily understood by users and facilitates access to information. It is not always the case that the display of names in alphabetical order can cause damage. The alphabetical list of candidates who have passed a job contest is harmless. However, if the same list is used to support a candidate selection in a recruitment process, it cannot be guaranteed to be completely harmless. In fact, the conclusion previously mentioned, had already been recognized in a more general context of the lexicographic order. \cite{fishburn1974exceptional} makes reference to the behavioral current represented by studies on rational choices \cite{simon1955behavioral} that the lexicographic order presupposes the element's choice preference based on the order of presentation. As the alphabetical order is a type of lexicographic order, the way information is presented in the interfaces ends up becoming an additive effect in this choice process that privileges the first elements. 

The results presented in this article show that for certain situations, especially those where people's names are listed for the purpose of supporting choice among the listed persons, the alphabetical order is not harmless and should draw the attention of software engineers and IHC designers. The development of tools and guidelines to be followed by these professionals can help them both in investigating the risk that a certain list of alphabetically ordered names may contain, as well as mitigating the negative effects that can be seen by viewing this list.

\section{Related Work}

To the best of our knowledge this work is the first work  to empirically investigate the potential impact of names  that are alphabetically ordered on models of preferences and choices in decision making. Depending on the type of the decision making model, lexicographically ordered names could induce biased choices. The characterization of the  display of these names for user selection via scrolling has also been done. 

Other works focus on measuring fairness in classified lists \cite{yang2017measuring,zehlike2017fa}, analyzing lists that meet the criteria of fairness \cite{carney2012first} by selecting fair and diverse sets \cite{stoyanovich2018online}. Yang and Stoyanovich \cite{yang2017measuring} studied the problem of fairness in rankings. Like us they propose a statistical parity measure based on comparing the distributions of a list of individuals in different prefixes of the list (for example {\it Top-10, Top-20, Top-30}) and calculate the mean of these differences with a logarithmic discount \cite{jarvelin2002cumulated}. It is assumed that those who come first in a ranking benefit the most. This stems from the underlying assumption that biased decisions always provide a positive result (bonus, selective choice) to the individuals on the list. 

There is a large amount of work studying the impact of the order of names in the context of scientific publications. Haque and Ginsparg \cite{haque2009positional} noted that the position of the article in the list of announcements of ArXiv is correlated with the citations of this article. Feenberg, Ganguli, Gaule, and Gruber \cite{feenberg2017s} have shown that there are similar biases for downloads and citations of papers from NBER "New This Week" working papers. Ray and Robison \cite{ray2018certified} also suggest a new alternative citation mechanism (certified random order), in place of the alphabetical order of scientific articles as traditionally has been done in research in Economics \cite{jacobs2015alphabetic}. Einav and Yariv \cite{einav2006s} argue that several criteria for the success of researchers and professors in the US economy (tenure at highly ranked schools, the Nobel Prize and Clark Medal awards) are correlated with the initial letter of the surname, favoring those with letters at the beginning in the alphabet, and this continues to occur when categorized by country, ethnicity and religion. \cite{weber2018effects} in a recent review, show several other examples and concludes that there is convinving evidence that alphabetical discrimination exists and that researchers react to it. Marketing research suggests that products presented first, have a higher probability of selection  \cite{carney2012first}. In the context of the stock exchange, shares with names that begin with initial letters of the alphabet are more likely to be traded \cite{jacobs2015alphabetic}.

\section{Conclusion}

The results presented in this research shed light on a problem that had not been recognized yet by researchers in Computer and Information Science. Making recommendations that consist of good programming and interface practice guidelines can support developers in circumstances where alphabetical ordering of names and the consequent display of those names on devices may induce biased choices. For legacy systems, it also opens the way to understand the magnitude of the problem from the analysis of software already developed and that may be producing alphabetical orders that show implicit ordering that causes an imbalance in the population viewed. The number of examples in proprietary information systems is possibly yet much larger than systems publicly open on the Web.

Solutions to mitigate the problems described here are out of the scope of this article. Some avenues are using methods that combine ordering and equitable balancing of the samples to be viewed, similar to that proposed by \cite{stoyanovich2018online}.  Algorithms to achieve the equal balance have been recently proposed \cite{asudeh2017designing,celis2017ranking}. Another alternative is to follow what has been proposed by Zehlike et al. \cite{zehlike2017fa} who claims that the proportion of members of a protected group(such as gender, ethnic majority/minority,or disability status) in each classification prefix remains statistically above a given threshold. Celis et al. \cite{carney2012first} provide a theoretical investigation of classification with restrictions of fairness. In their work, impartiality in a classification list is quantified as a limit superior to the number of items in the {\it Top-k} that belong to several, possibly overlapping, types. However, all these alternative solutions requires usability tests for measuring the impact on the user. 

Besides the mentioned impacts, new lines of research can investigate how software engineering practices can avoid biased results. Generalization can occur either at the level of the entity’s name being ordered (e.g., company names, place names) or at the level of the attribute to be evaluated. In this article we focus on gender, but there are other properties, such as race, that may behave similarly, since people’s names are usually associated with ethnic factors. For race, an analysis by family name is more appropriate. A temporal analysis of the names the most frequent during the years may also shed light on how much the harm of the alphabetical order of names varies over time. Another potential avenue might be nationality bias in alphabetical ranking by surname. For example, are Scandinavian or Korean names disadvantaged by alphabetical ordering by surname when a multinational dataset of names is used?


\begin{thebibliography}{10}

\bibitem{friedman1996bias}
Batya Friedman and Helen Nissenbaum.
\newblock Bias in computer systems.
\newblock {\em ACM Transactions on Information Systems (TOIS)}, 14(3):330--347,
  1996.

\bibitem{shifrin1985justice}
CA~Shifrin.
\newblock Justice will weigh suit challenging airlines computer reservations.
\newblock {\em AVIATION WEEK \& SPACE TECHNOLOGY}, 122(12):105, 1985.

\bibitem{mohd1990loophole}
I~MOHD~TAIB.
\newblock Loophole allows bias in displays on computer reservations systems.
\newblock {\em Aviation Week \& Space Technology}, 132(7), 1990.

\bibitem{carney2012first}
Dana~R Carney and Mahzarin~R Banaji.
\newblock First is best.
\newblock {\em PloS one}, 7(6):e35088, 2012.

\bibitem{einav2006s}
Liran Einav and Leeat Yariv.
\newblock What's in a surname? the effects of surname initials on academic
  success.
\newblock {\em Journal of Economic Perspectives}, 20(1):175--187, 2006.

\bibitem{feenberg2017s}
Daniel Feenberg, Ina Ganguli, Patrick Gaule, and Jonathan Gruber.
\newblock It’s good to be first: Order bias in reading and citing nber
  working papers.
\newblock {\em Review of Economics and Statistics}, 99(1):32--39, 2017.

\bibitem{haque2009positional}
Asif-ul Haque and Paul Ginsparg.
\newblock Positional effects on citation and readership in arxiv.
\newblock {\em Journal of the American Society for Information Science and
  Technology}, 60(11):2203--2218, 2009.

\bibitem{jacobs2015alphabetic}
Heiko Jacobs and Alexander Hillert.
\newblock Alphabetic bias, investor recognition, and trading behavior.
\newblock {\em Review of Finance}, 20(2):693--723, 2015.

\bibitem{zehlike2017fa}
Meike Zehlike, Francesco Bonchi, Carlos Castillo, Sara Hajian, Mohamed Megahed,
  and Ricardo Baeza-Yates.
\newblock Fa* ir: A fair top-k ranking algorithm.
\newblock In {\em Proceedings of the 2017 ACM on Conference on Information and
  Knowledge Management}, pages 1569--1578. ACM, 2017.

\bibitem{fitts1954information}
Paul~M Fitts.
\newblock The information capacity of the human motor system in controlling the
  amplitude of movement.
\newblock {\em Journal of experimental psychology}, 47(6):381, 1954.

\bibitem{pirolli2007information}
Peter Pirolli.
\newblock {\em Information foraging theory: Adaptive interaction with
  information}.
\newblock Oxford University Press, 2007.

\bibitem{adoption-system}
American adoptions.
\newblock
  Available:\url{https://www.americanadoptions.com/family_profile/browse?family_types\%5B3\%5D=Same+Sex+Male&family_types\%5B4\%5D=Same+Sex+Female&fpch=&search=search&embed=&onLoadScrollTo=fp-list},
  2019.
\newblock Accessed: 2019-08-15.

\bibitem{brazilian-system}
Busca de candidatos.
\newblock
  Available:\url{https://www1.folha.uol.com.br/poder/eleicoes/candidatos/\#2018/deputado-estadual-distrital/sp
  }, 2019.
\newblock Accessed: 2019-08-15.

\bibitem{hassenzahl2007hedonic}
Marc Hassenzahl.
\newblock The hedonic/pragmatic model of user experience.
\newblock {\em Towards a UX manifesto}, 10, 2007.

\bibitem{hassenzahl2015experience}
Marc Hassenzahl, Annika Wiklund-Engblom, Anette Bengs, Susanne H{\"a}gglund,
  and Sarah Diefenbach.
\newblock Experience-oriented and product-oriented evaluation: psychological
  need fulfillment, positive affect, and product perception.
\newblock {\em International journal of human-computer interaction},
  31(8):530--544, 2015.

\bibitem{hassenzahl2018thing}
Marc Hassenzahl.
\newblock The thing and i: understanding the relationship between user and
  product.
\newblock In {\em Funology 2}, pages 301--313. Springer, 2018.

\bibitem{culnan2000protecting}
Mary~J Culnan.
\newblock Protecting privacy online: Is self-regulation working?
\newblock {\em Journal of Public Policy \& Marketing}, 19(1):20--26, 2000.

\bibitem{data-franca}
Insee.
\newblock Available:https:\url{//www.insee.fr/fr/statistiques/3532172}, 2019.
\newblock Accessed: 2019-03-20.

\bibitem{data-usa}
Social security administration.
\newblock Available: \url{https://www.ssa.gov/oact/babynames/limits.html},
  2019.
\newblock Accessed: 2019-01-15.

\bibitem{info-pop-usa}
World bank group.
\newblock
  Available:\url{https://data.worldbank.org/country/united-states?view=chart},
  2019.
\newblock Accessed: 2019-05-22.

\bibitem{data-escocia}
National records of scotland.
\newblock Available:
  \url{https://www.nrscotland.gov.uk/statistics-and-data/statistics/statistics-by-theme/vital-events/names/babies-first-names/babies-first-names-summary-records-comma-separated-value-csv-format},
  2019.
\newblock Accessed: 2019-01-15.

\bibitem{data-esp}
National statistics institute.
\newblock Available: \url{http://www.ine.es}, 2019.
\newblock Accessed: 2019-01-15.

\bibitem{saeed2014fisher}
Swaleha Saeed, M~Sarosh Umar, M~Athar Ali, and Musheer Ahmad.
\newblock Fisher-yates chaotic shuffling based image encryption.
\newblock {\em arXiv preprint arXiv:1410.7540}, 2014.

\bibitem{yang2017measuring}
Ke~Yang and Julia Stoyanovich.
\newblock Measuring fairness in ranked outputs.
\newblock In {\em Proceedings of the 29th International Conference on
  Scientific and Statistical Database Management}, page~22. ACM, 2017.

\bibitem{nadaraya1964}
Elizbar~A Nadaraya.
\newblock On estimating regression.
\newblock {\em Theory of Probability \& Its Applications}, 9(1):141--142, 1964.

\bibitem{watson1964}
Geoffrey~S Watson.
\newblock Smooth regression analysis.
\newblock {\em Sankhy{\=a}: The Indian Journal of Statistics, Series A}, pages
  359--372, 1964.

\bibitem{racine2004}
Jeff Racine and Qi~Li.
\newblock Nonparametric estimation of regression functions with both
  categorical and continuous data.
\newblock {\em Journal of Econometrics}, 119(1):99--130, 2004.

\bibitem{nielsen1999designing}
Jakob Nielsen.
\newblock {\em Designing web usability: The practice of simplicity}.
\newblock New riders publishing, 1999.

\bibitem{pemberton2003hotel}
Steven Pemberton.
\newblock Hotel heartbreak.
\newblock {\em interactions}, 10(5):64, 2003.

\bibitem{mackenzie1992fitts}
I~Scott MacKenzie.
\newblock Fitts' law as a research and design tool in human-computer
  interaction.
\newblock {\em Human-computer interaction}, 7(1):91--139, 1992.

\bibitem{fishburn1974exceptional}
Peter~C Fishburn.
\newblock Exceptional paper—lexicographic orders, utilities and decision
  rules: A survey.
\newblock {\em Management science}, 20(11):1442--1471, 1974.

\bibitem{simon1955behavioral}
Herbert~A Simon.
\newblock A behavioral model of rational choice.
\newblock {\em The quarterly journal of economics}, 69(1):99--118, 1955.

\bibitem{stoyanovich2018online}
Julia Stoyanovich, Ke~Yang, and HV~Jagadish.
\newblock Online set selection with fairness and diversity constraints.
\newblock In {\em Proceedings of the EDBT Conference}, 2018.

\bibitem{jarvelin2002cumulated}
Kalervo J{\"a}rvelin and Jaana Kek{\"a}l{\"a}inen.
\newblock Cumulated gain-based evaluation of ir techniques.
\newblock {\em ACM Transactions on Information Systems (TOIS)}, 20(4):422--446,
  2002.

\bibitem{ray2018certified}
Debraj Ray and Arthur Robson.
\newblock Certified random: A new order for coauthorship.
\newblock {\em American Economic Review}, 108(2):489--520, 2018.

\bibitem{weber2018effects}
Matthias Weber.
\newblock The effects of listing authors in alphabetical order: a review of the
  empirical evidence.
\newblock {\em Research Evaluation}, 27(3):238--245, 2018.

\bibitem{asudeh2017designing}
Abolfazl Asudeh, HV~Jagadish, Julia Stoyanovich, and Gautam Das.
\newblock Designing fair ranking schemes.
\newblock {\em arXiv preprint arXiv:1712.09752}, 2017.

\bibitem{celis2017ranking}
L~Elisa Celis, Damian Straszak, and Nisheeth~K Vishnoi.
\newblock Ranking with fairness constraints. corr abs/1704.06840 (2017).
\newblock {\em arXiv preprint arXiv:1704.06840}, 2017.

\end{thebibliography}
\end{document}